\input{aipcheck}

\documentclass[
    ,final            
  ]
  {aipproc}

\layoutstyle{8x11single}

\begin{document}

\title{The Status of Neutralino Dark Matter}


\author{Bibhushan Shakya}{
  address={Laboratory for Elementary Particle Physics,\\
  Cornell University, Ithaca, NY 14853, USA}
}

\begin{abstract}
The lightest neutralino in supersymmetry is the most studied dark matter candidate. This writeup reviews the status of neutralino dark matter in minimal and nonminimal supersymmetric models in light of recent null results at the XENON100 experiment and the observation of a 130 GeV gamma ray signal from the Galactic Center by the Fermi LAT. 
 \end{abstract}

\maketitle


\section{Introduction}

Although the identity of dark matter remains a mystery, the lightest neutralino in supersymmetry, when also the lightest supersymmetric particle (LSP) protected by R-parity, is the most studied dark matter candidate. The possible experimental signatures of the neutralino have been studied extensively, and parts of the most reasonable parameter space in supersymmetry predict positive signals in one or more of the current experiments. There are three traditional approaches to studying dark matter interactions: direct production at colliders, direct detection of dark matter scattering off atomic nuclei, and indirect detection of dark matter decay or annihilation products in the solar neighborhood. In the past few years, experiments on all three fronts have begun to probe regions of parameter space where supersymmetric dark matter predicts positive signals, hence the picture of supersymmetric dark matter has been changing rapidly. 

On the collider front, the Large Hadron Collider (LHC) has found no missing energy signatures predicted from dark matter; however, it has unearthed a 126 GeV Standard-Model-like Higgs boson, which has strong implications for minimal and nonminimal models of supersymmetry \cite{HPR}. Among direct detection experiments, the best upper bounds on the cross section of elastic, spin-independent dark matter-nucleon scattering in the 10 GeV - 1 TeV mass range come from the XENON100 experiment~\cite{xenon100paper,xenon2012}. There has been greater excitement on the indirect detection front, where a sharp feature at an energy of approximately 130\,GeV in the gamma ray spectrum towards the Galactic Center in the data gathered by the Fermi Large Area Telescope (LAT) \cite{130weniger,130fermi} holds the possibility of having its origin in dark matter. This writeup discusses the implications of these experimental results on the viability of neutralino dark matter. It should be pointed out that there exist many other strong constraints on supersymmetric dark matter that will not be covered in this writeup.

\section{The Fermi 130 GeV Signal in Supersymmetry}

Several analyses \cite{130fermi} have recently confirmed the presence of a sharp feature at an energy of approximately 130\,GeV in the gamma ray spectrum towards the Galactic Center in the data gathered by the Fermi Large Area Telescope (LAT). Although the significance of this signal has since then decreased, it is still of considerable interest, since such a feature has long been earmarked as a ``smoking gun" signature of dark matter annihilation in the galaxy. 

Assuming a dark matter origin, the signal is best fit by a 130\,GeV dark matter particle pair-annihilating into photons with an annihilation cross section of $\langle\sigma v\rangle_{\gamma\gamma}=1.27\times10^{-27}$cm$^3$s$^{-1}$, assuming an Einasto profile for the dark matter distribution \cite{130weniger}. Since dark matter is not expected to couple directly to photons, annihilation to a photon pair must occur via a loop; for a thermal relic, this loop-suppressed cross section is generally too small to produce the signal observed by  Fermi. Second, even if this cross section can be made large enough, tree-level annihilation to particles that mediate the photon pair production process should produce a large continuum of photons at lower energies, which is not seen in the  Fermi data. These considerations have been shown to rule out neutralino dark matter as an explanation of this line signal \cite{continuum, continuum2}.  

A monochromatic line signal, however, is not the only possibility that can explain this feature; it is well-known that internal bremsstrahlung (hereafter IB) --  the production of a photon in conjunction with the leading annihilation channel into charged particles -- can also give sharp spectral features in the $\gamma$ ray spectrum close to the dark matter mass \cite{ib}. This section will explore this possibility, and is based on \cite{130susy} (see also \cite{tcreview} for a similar study).

\subsection{Internal Bremsstrahlung from Neutralino Dark Matter}

The IB component from neutralino annihilation is known to be the most prominent when annihilation is into particles that are effectively massless relative to the neutralino, and the virtual particle that mediates the process is close in mass to the neutralino \cite{ib}. Since the W and Z gauge bosons are massive final states for a neutralino of mass around 130-150\,GeV, IB from wino or Higgsino dark matter does not produce a feature sharp enough to explain the Fermi observation, despite the presence of a degenerate chargino to mediate its annihilation; this has been verified explicitly.  

IB from bino dark matter, on the other hand, is more promising. The main annihilation channels for a bino are to fermion pairs, mediated by the corresponding sfermions. For nonrelativistic annihilation in the halo, the cross section for this process is helicity suppressed by a factor of $(m_f/m_{\chi})^2$; since the top quark is heavier than the dark matter mass of interest here and all other SM fermions are $\mathcal{O}$(GeV) or lighter, this suppression is of several orders of magnitude, and acts as an efficient mechanism to suppress the continuum photon production. The addition of a photon in the final state, on the other hand, lifts this helicity suppression, and $\sigma v(\chi\chi\rightarrow\bar{f}f\gamma)$ can be comparable to $\sigma v(\chi\chi\rightarrow\bar{f}f)$. Since fermions and sfermions couple to the bino via hypercharge and leptons have larger hypercharge than quarks (also, sleptons are generally lighter than squarks),  IB primarily involves leptonic channels, with light sleptons a crucial element to enhance the signal.

\begin{figure}[h]
\centering
\includegraphics[width=3.0in,height=2.1in]{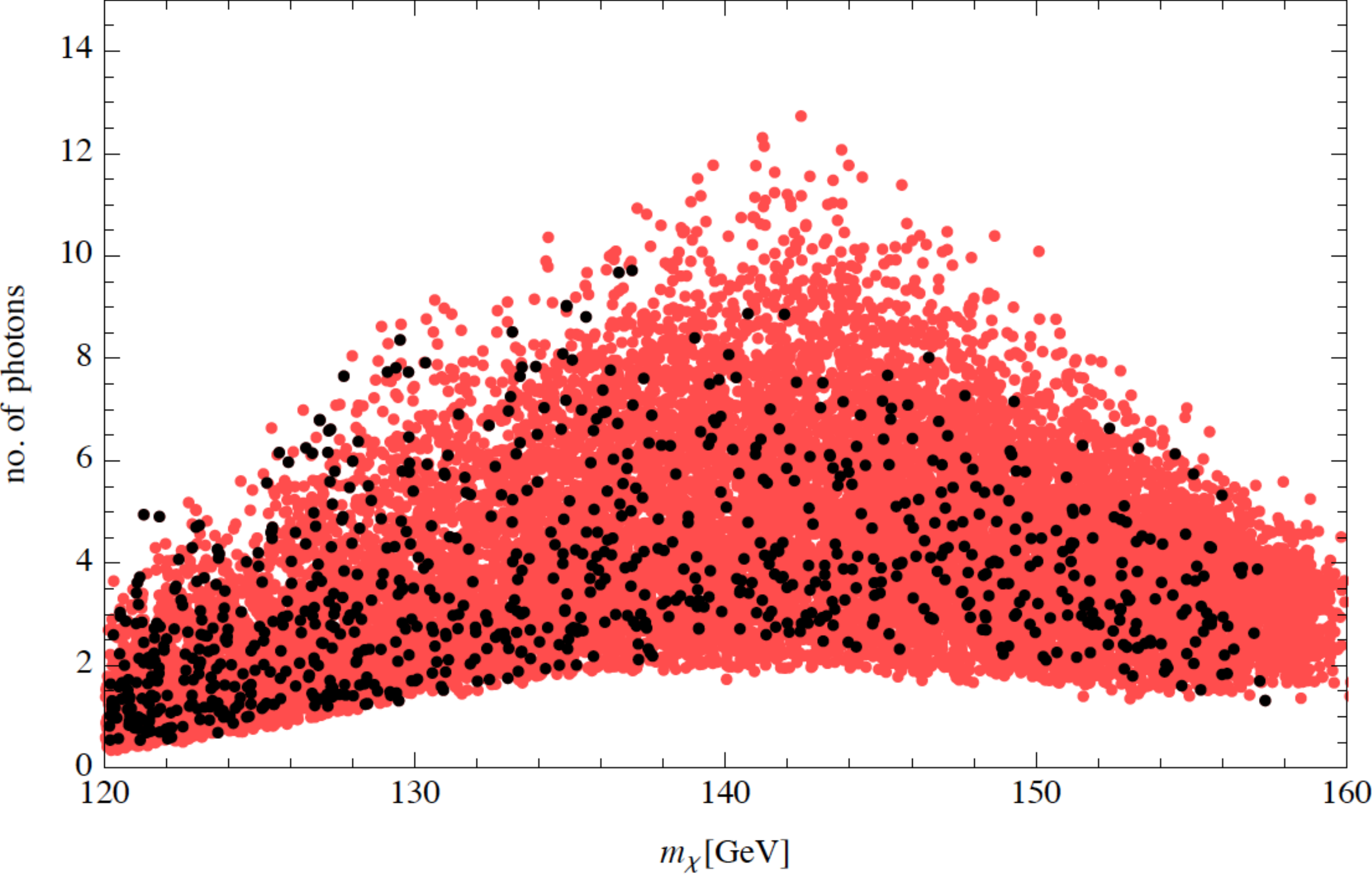}
\includegraphics[width=3.0in,height=2.1in]{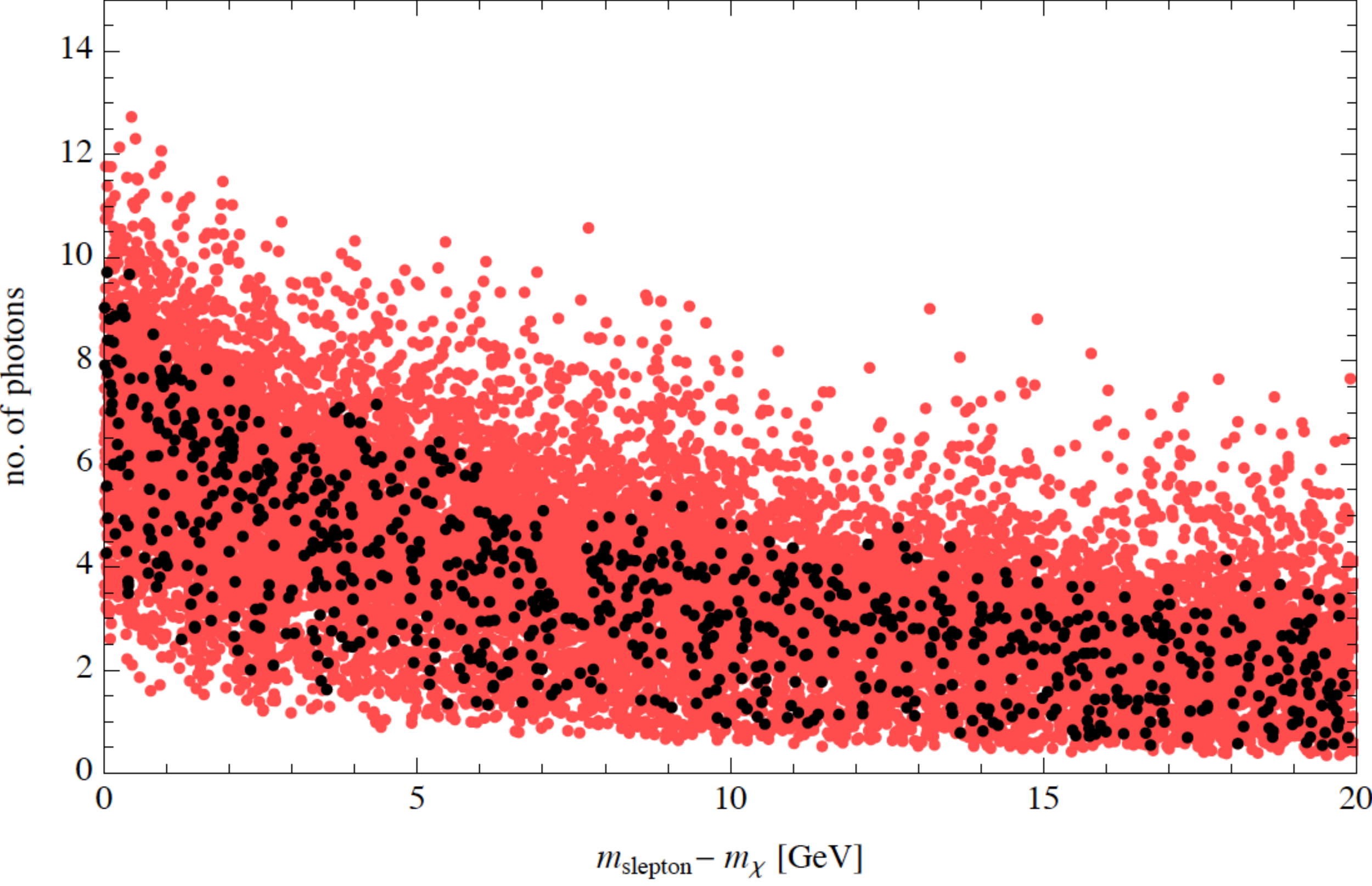}
\caption{The total number of photons in the 121.62\,GeV to 136.40\,GeV energy range from dark matter annihilation, as a function of dark matter mass (left) and slepton-neutralino mass difference (right), for a scan over primarily bino candidates. Points with thermal relic density calculated to be in the range $0.1\leq \Omega h^2\leq 0.124$, making them consistent with relic density constraints, are shown in black. }
\label{fig:photonnumber}
\end{figure}

Figure \ref{fig:photonnumber} shows the total number of photons from IB in the signal region, and the dependence on the slepton-neutralino mass difference from a scan over bino dark matter and light sleptons\footnote{In addition to the IB, the Inverse Compton Scattering contribution from dark matter annihilation was estimated using a semi-analytic formalism described in \cite{positrons}, and was found to be negligible.}. The interested reader is referred to \cite{130susy} for details of the scan. The figure [left] shows that dark matter contributes $\mathcal{O}$(few) photons, peaking at $\sim$ 13 photons around $m_\chi\approx 145\,$GeV, in the given energy range (the Fermi data set contains 24 photons in this region). The occurrence of the peak at this energy is understandable: for $m_\chi\approx145$\,GeV, the $\gamma Z$ channel gives monoenergetic photons at $E_\gamma=m_\chi (1-m_Z^2/4m_\chi^2)\approx130$\,GeV, which can be sizable even with a tiny wino/Higgsino component. Meanwhile, the peak of the IB feature still falls mostly into the 121.62\,GeV to 136.40\,GeV range (recall that the best fit for a purely IB signal occurs for $m_\chi\sim150$\,GeV); hence both IB and the $\gamma Z$ line are at ideal energies to contribute towards the signal. As the mass changes away from this ideal value, either the IB peak or the line contribution is lost, and the signal dies down. The figure on the right confirms that the photon count gradually rises as the sleptons become more degenerate in mass with the neutralino, as is expected of IB.

\subsection{Benchmark Points and Fit to Data}

\begin{figure}[h]
\centering
\begin{tabular}{cc}
\includegraphics[width=2.9in,height=1.8in]{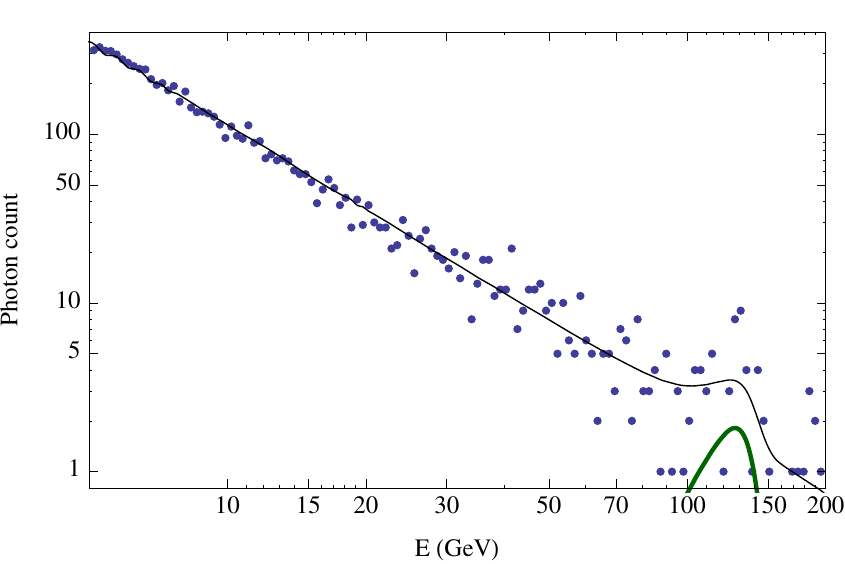} &
\includegraphics[width=2.9in,height=1.8in]{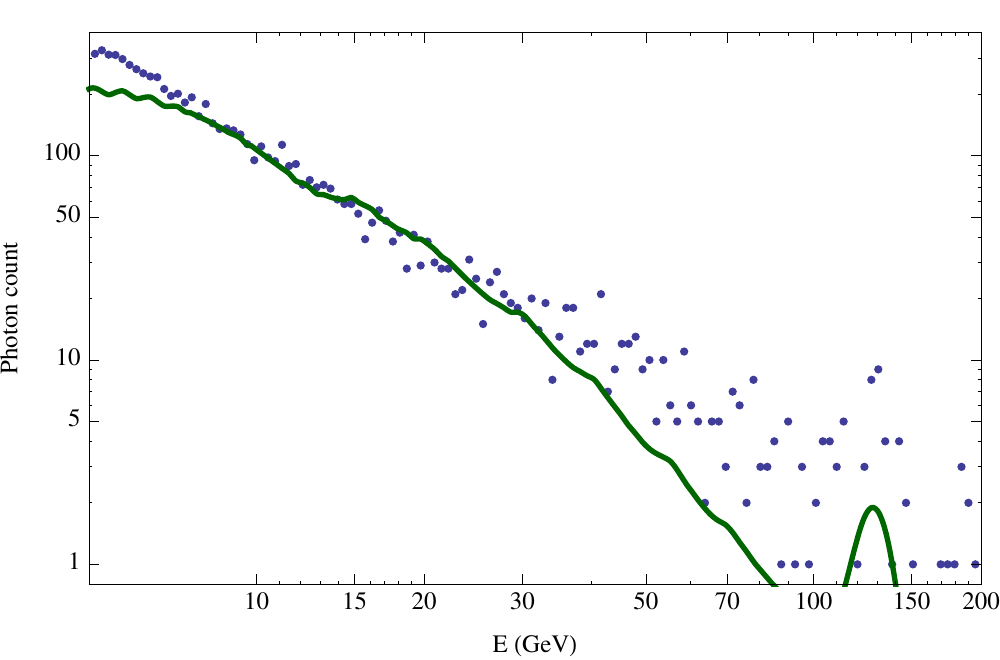}\\
\includegraphics[width=2.9in,height=1.8in]{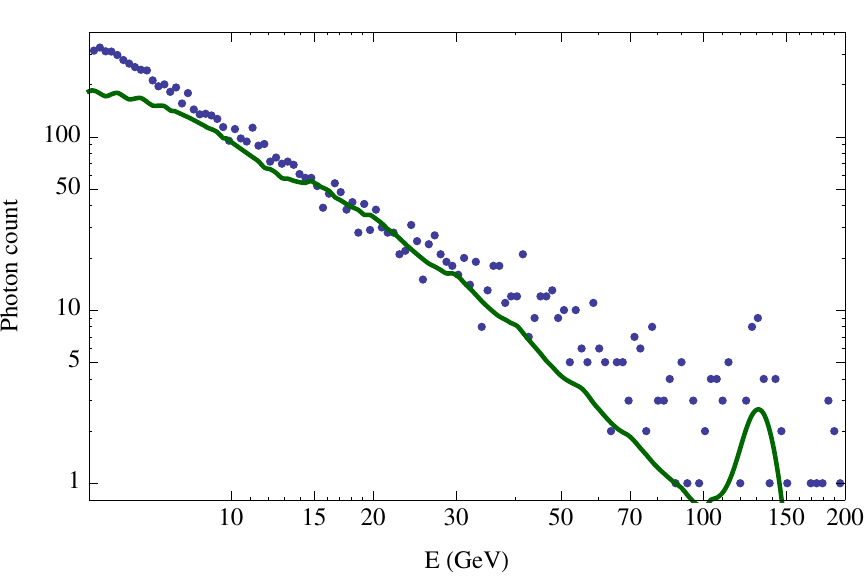} &
\includegraphics[width=2.9in,height=1.8in]{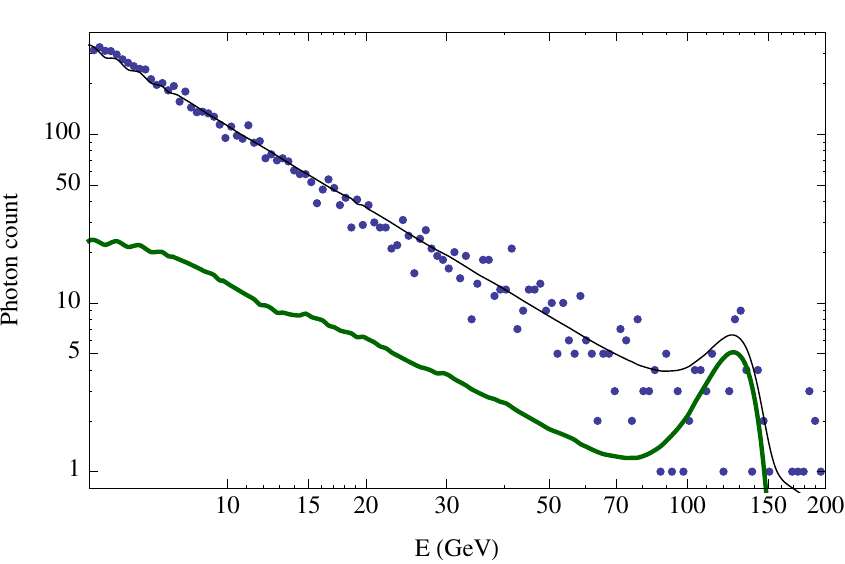}\\
\end{tabular}
\caption{Fermi data from the inner 3$^{\circ}$ of the Galactic Center  for all 128 energy bins (blue dots, as listed in Appendix A of \cite{continuum}) and the gamma ray spectra from dark matter (green) for the four benchmark points BM1, BM2 (top row), BM3, and BM4 (bottom row). The black curves for BM1 and BM4 represent the overall fit to the signal, consisting of a single power law background in addition to the dark matter signal; BM2 and BM3 saturate the continuum at lower energies and do not allow such fits.
}
\label{fig:fits1}
\end{figure}

This section discusses four benchmark points (labelled BM1, BM2, BM3, and BM4) that are representative of the scanned sample, and their fits to the Fermi signal. These points are listed in Table \ref{tab:bm}, with other relevant information, while the resultant fits are shown in Figure \ref{fig:fits1}. The interested reader is referred to \cite{130susy} for details of the fit procedure.

\begin{table}[h]
\centering
\begin{tabular}{|c|c|c|c|c|} \hline
~ & ~~~$BM1$~~~&~~~$BM2$~~~ & ~~~$BM3$~~~&~~~$BM4$~~~\\
\hline
$M_1$ & 135.2 & 144.7  &145.6 & 138.2\\
$M_2$& 235.5 & 152.8 &150.4 & 161.2\\
$\mu$ & -489.9 & 838.4 & 783.0 & 512.9\\
tan$\beta$& 18.5 & 6.6 & 33.2 & 20.5\\
$m_{\tilde{l}}$& 136.7 & 156.6 & 146.7 & 138.5\\
\hline
$m_\chi$ & 134.4 & 143.0 & 144.7 & 136.4\\
bino fraction & 0.99 & .90 & 0.91 & 0.97\\
$\Omega h^2$ & 0.19 & .0058 & 0.0033 & 0.11\\
$n_\gamma$ from IB & 4.8 & 1.8 & 4.5 & 14.7\\
$n_\gamma$ ($\gamma\gamma +\gamma Z$) & 2.0 & 5.1 & 5.2 & 4.4\\
TS & 15.8 & -- & -- & 17.8\\
significance & 4.0 & -- & -- & 4.2\\
\hline
\end{tabular}
\caption{Four benchmark points chosen for detailed study and fit to Fermi data, and fit results. All masses are in GeV. $n_\gamma$ refers to the number of photons contributed to the 121.62\,GeV to 136.40\,GeV energy bin.}
\label{tab:bm}
\end{table}

BM1 is an almost pure bino that contributes dominantly via IB, with the continuum almost nonexistent and the almost degenrate sleptons making the IB component very prominent. BM2 improves on BM1 by introducing a 10\% wino admixture to the neutralino, opening up significant contributions from the $\gamma Z$ line, but at the cost of continuum photons. BM3 combines the virtues of both BM1 and BM2: sleptons degenerate with the neutralino lead to a sharp IB feature, while a small wino admixture contributes a prominent $\gamma Z$ line signal, resulting in a relative abundance of photons in the right energy bin. BM4 represents a fit achieved by enhancing the dark matter signal by an additional factor of 3, and could correspond to some astrophysical enhancement such as a steeper dark matter profile at the Galactic Center or substructure along the line of sight towards this region, resulting in a significantly better fit. It should be stressed that these were chosen to highlight distinct features of signals that are possible with internal bremsstrahlung, and are not the points that best fit the data. 

\section{Naturalness Implications of XENON100 Results}

Next, we turn to the theoretical implications of the null XENON100 results\cite{xenon100paper,xenon2012} on various models of supersymmetry. While the minimal supersymmetric standard model (MSSM) is the simplest and most studied version, the LHC discovery of a 126 GeV Higgs is more naturally realized in the Next-to-Minimal Supersymmetric Standard Model (NMSSM), and its variant, $\lambda$-SUSY \cite{HPR}; hence we focus on these three models in this section. In particular, direct detection cross-sections are studied in the context of the amount of fine-tuning in electroweak symmetry breaking (EWSB). Details of how this fine-tuning is calculated in various models is described in \cite{ftmssm, ftsusy2}; for the purpose of this writeup, it suffices to note that large values of the $\mu$ parameter correspond to greater fine-tuning in EWSB.

\subsection{MSSM}

\begin{figure}[h]
\centering
\includegraphics[width=3.0in,height=2.1in]{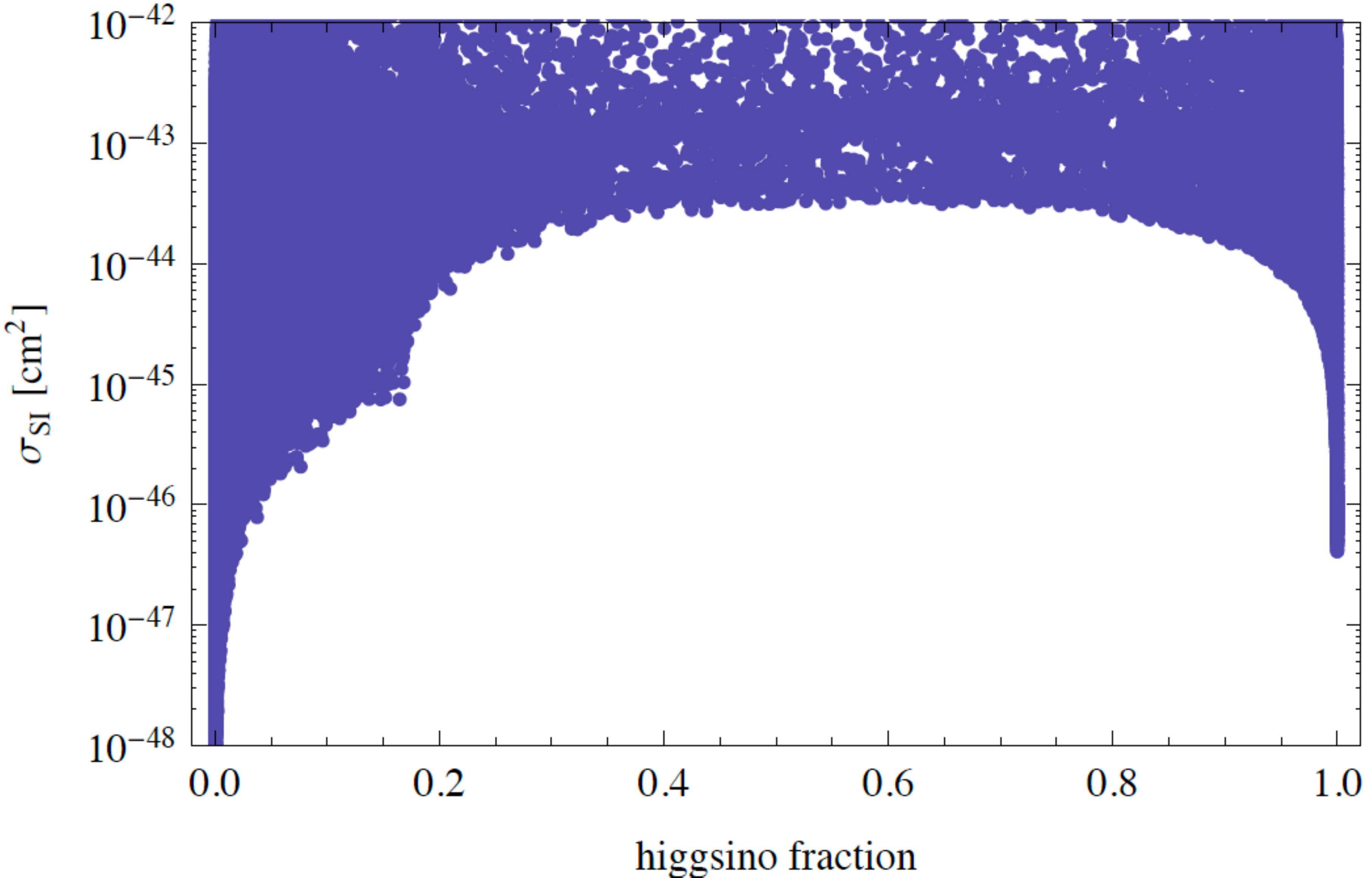}
\includegraphics[width=3.0in,height=2.1in]{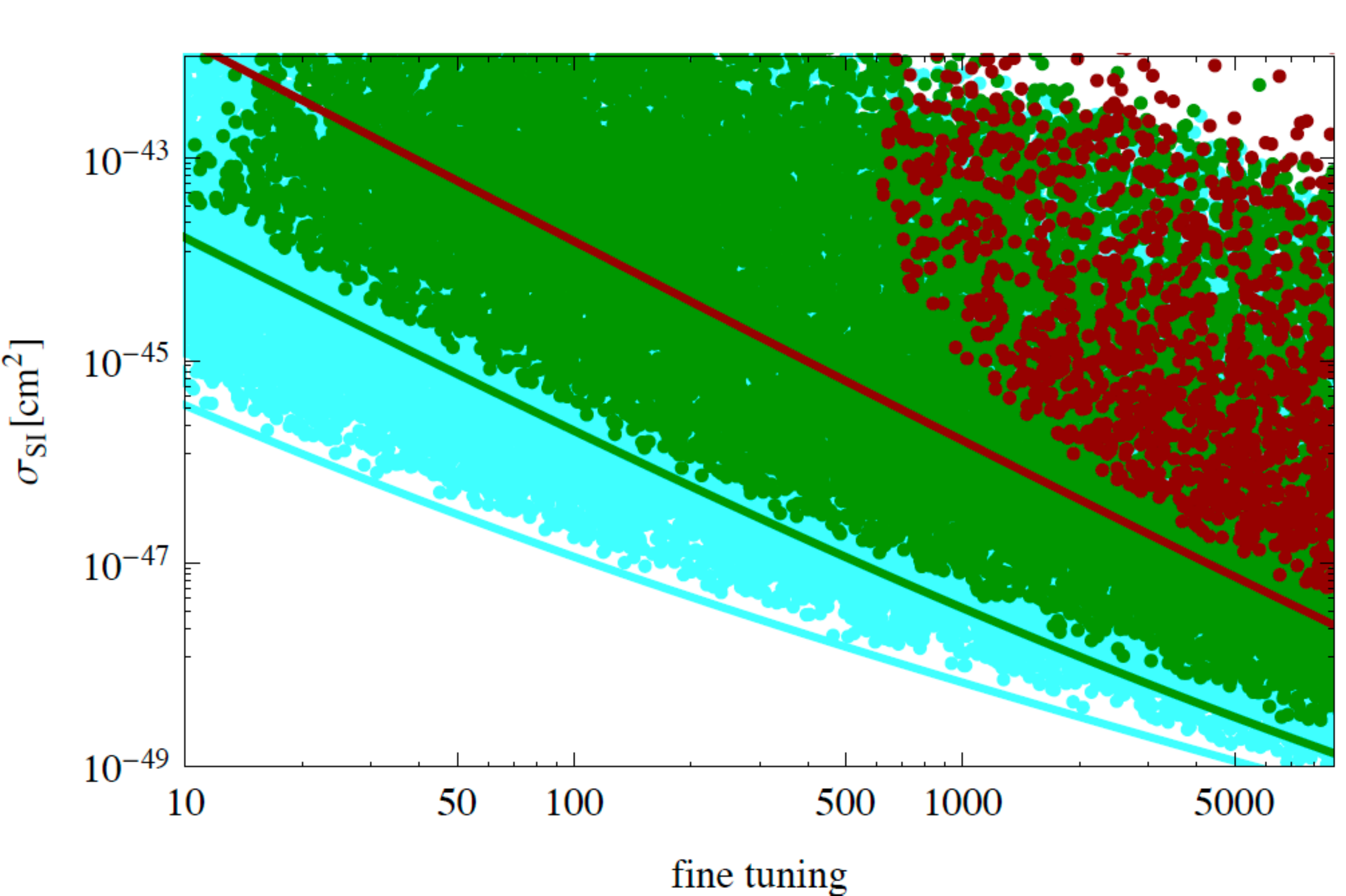}
\caption{Left: Direct detection cross section vs. Higgsino fraction of the neutralino. Right: Direct detection cross section vs. EWSB fine-tuning, for gaugino-like neutralino. Cyan, green and red points correspond to the dark matter particle mass in the intervals $[10, 100)$, $[100, 1000)$, and $[1000, 10^4]$ GeV, respectively. The cyan, green and red lines show the analytic lower bound (see \cite{ftmssm}) with $M_{ LSP}=10, 100, 1000$ GeV, respectively. Real, positive values of the MSSM parameters are assumed.}
\label{fig:hfraction}
\end{figure}

Ignoring squark mediated processes, which are likely subdominant, the direct detection cross section of neutralinos scattering off nuclei in the MSSM is mediated by the two CP-even Higgs bosons h and H. In the mass basis, the neutralino-neutralino-Higgs couplings have the form 
\begin{eqnarray}
\tilde{\chi}^0\tilde{\chi}^0h:& &~(gZ_{\chi2}-g^\prime Z_{\chi1})(\cos\alpha Z_{\chi4}+\sin\alpha Z_{\chi3} )\\
\tilde{\chi}^0\tilde{\chi}^0H:& &~(gZ_{\chi2}-g^\prime Z_{\chi1})(\sin\alpha Z_{\chi4}-\cos\alpha Z_{\chi3} )\,.
\label{nnh}
\end{eqnarray}
where $ \tilde{\chi}^0_1 = Z_{\chi 1} \tilde{B} + Z_{\chi 2} \tilde{W}^3 + Z_{\chi 3} \tilde{H}^0_d + Z_{\chi 4} \tilde{H}$.
If the $\tilde{\chi}^0\tilde{\chi}^0h$ is of its natural size ({\it i.e.} no accidental cancellations or small mixing angles are present), the direct detection cross section from $t$-channel Higgs exchange is of order (a few)$\times 10^{-44}$ cm$^2$ or above. Barring accidental cancellations, the only way to obtain a smaller cross section is to suppress this coupling by choosing the LSP to be an almost pure gaugino or Higgsino, which is precisely what is seen in Fig.~\ref{fig:hfraction} [left]. The interested reader is referred to \cite{ftmssm} for details of the scan used to produce the plots in this subsection. 

It is well-known that a pure Higgsino dark matter candidate needs to be at the TeV scale to have the correct relic density \cite{welltemp}. This requires the $\mu$ parameter to be at the TeV scale, introducing sub-percent level fine-tuning in EWSB. In the opposite regime, that of gaugino-like neutralino, the cross section can only be reduced by reducing the Higgsino admixture in the LSP, and the only way to achieve this is to again raise $\mu$. Therefore lower cross-sections must be correlated with greater fine-tuning in EWSB, which is seen in Fig.~\ref{fig:hfraction} [right] (see also \cite{ftmssm2} for similar studies).

\begin{figure}[th]
\centering
\includegraphics[width=3.0in,height=2.1in]{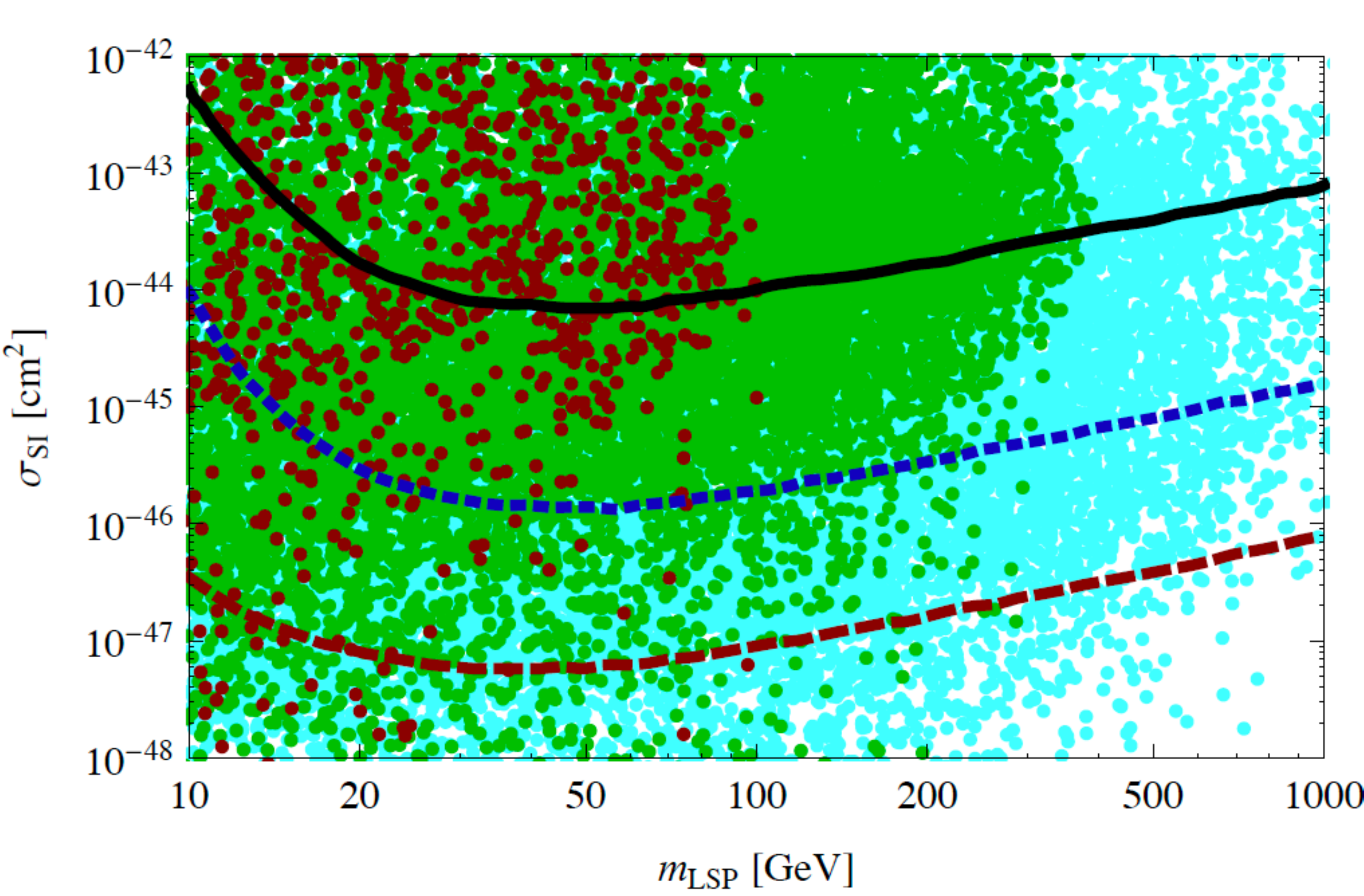}
\includegraphics[width=3.0in,height=2.1in]{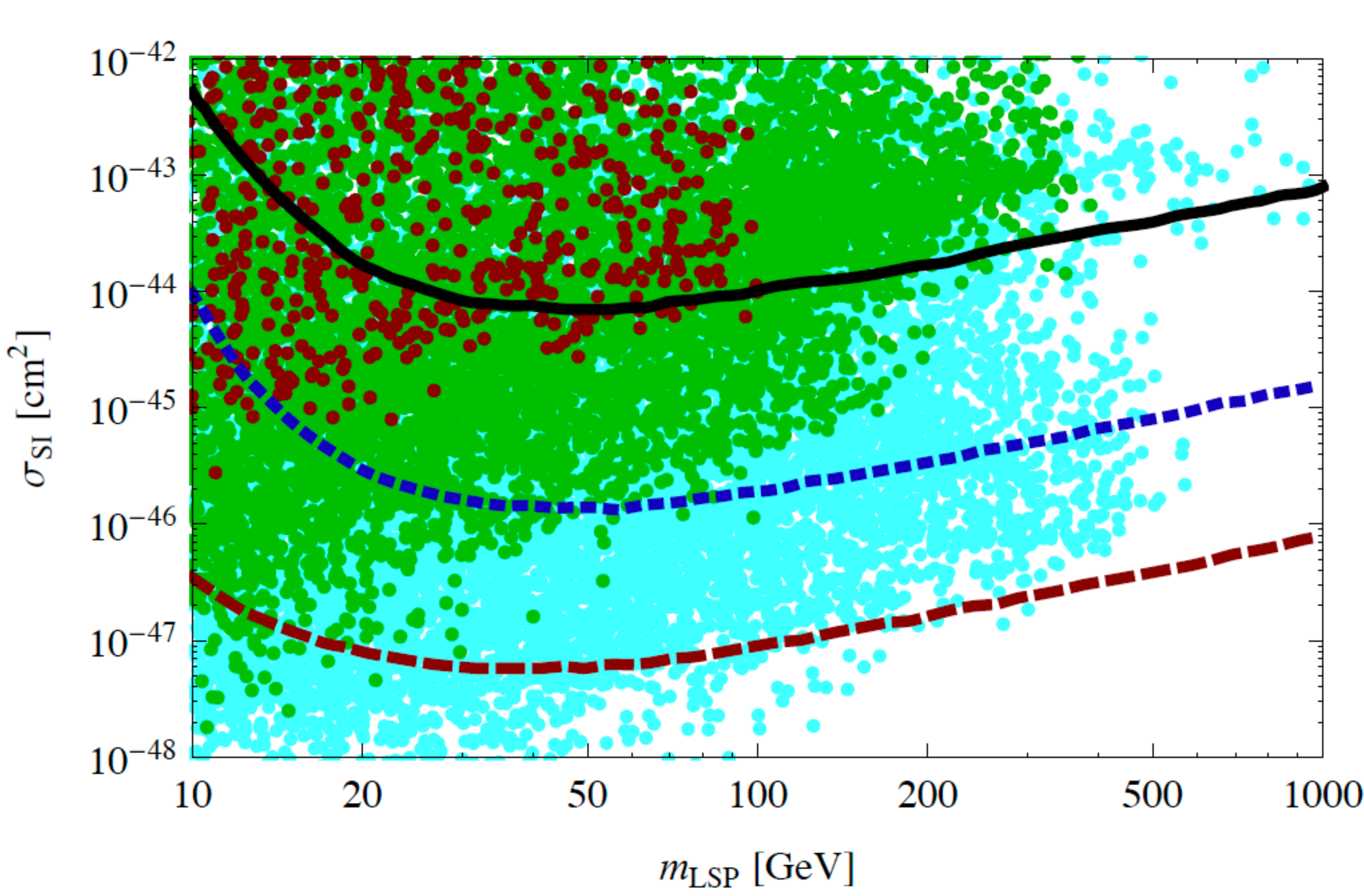}
\caption{Left panel: Direct detection cross section vs. dark matter particle mass, for gaugino-like neutralino.
Red, green and cyan points correspond to EWSB fine-tuning in the intervals $(0, 10)$; $[10, 100)$; and$[100,1000]$, respectively. Right panel: Same, with points with accidental cancellations in cross section removed. Current and projected XENON bounds (top to bottom: XENON100(2011), XENON100(2012) (earlier projection), XENON1T) are superimposed.}
\label{fig:MLSP_tuningALL}
\end{figure}

When the parameters are allowed to take on negative or complex values, the cross section can be small due to accidental cancellations among different terms in the cross section, which destroys the correlation between cross section and EWSB fine-tuning seen in Fig.~\ref{fig:hfraction} [right]. However, such cancellations are themselves fine-tuned (see \cite{ftmssm}), and removing points with such accidental cancellations again restores the correlation. The direct detection cross section vs. dark matter mass with and without points with such accidental cancellations is shown in \ref{fig:MLSP_tuningALL}, with various degrees of EWSB fine-tuning color-coded. 

\subsection {NMSSM and $\lambda$-SUSY}

Compared to the MSSM, the NMSSM and $\lambda$-SUSY contain an additional singlet superfield. The fermionic component of this superfield contributes an additional neutralino, the singlino, that can mix with the four MSSM neutralinos, while the scalar can mix with the CP-odd Higgses and provides an additional mediator for direct detection cross sections. In addition, to make a 126 GeV Higgs natural in these models, a large singlet-doublet superpotential coupling $\lambda$ is required. While $\lambda\textless 0.75$ in the NMSSM, it can be as large as 2 in $\lambda$-SUSY. 

\begin{figure}[t]
\centering
\includegraphics[width=3.0in,height=2.1in]{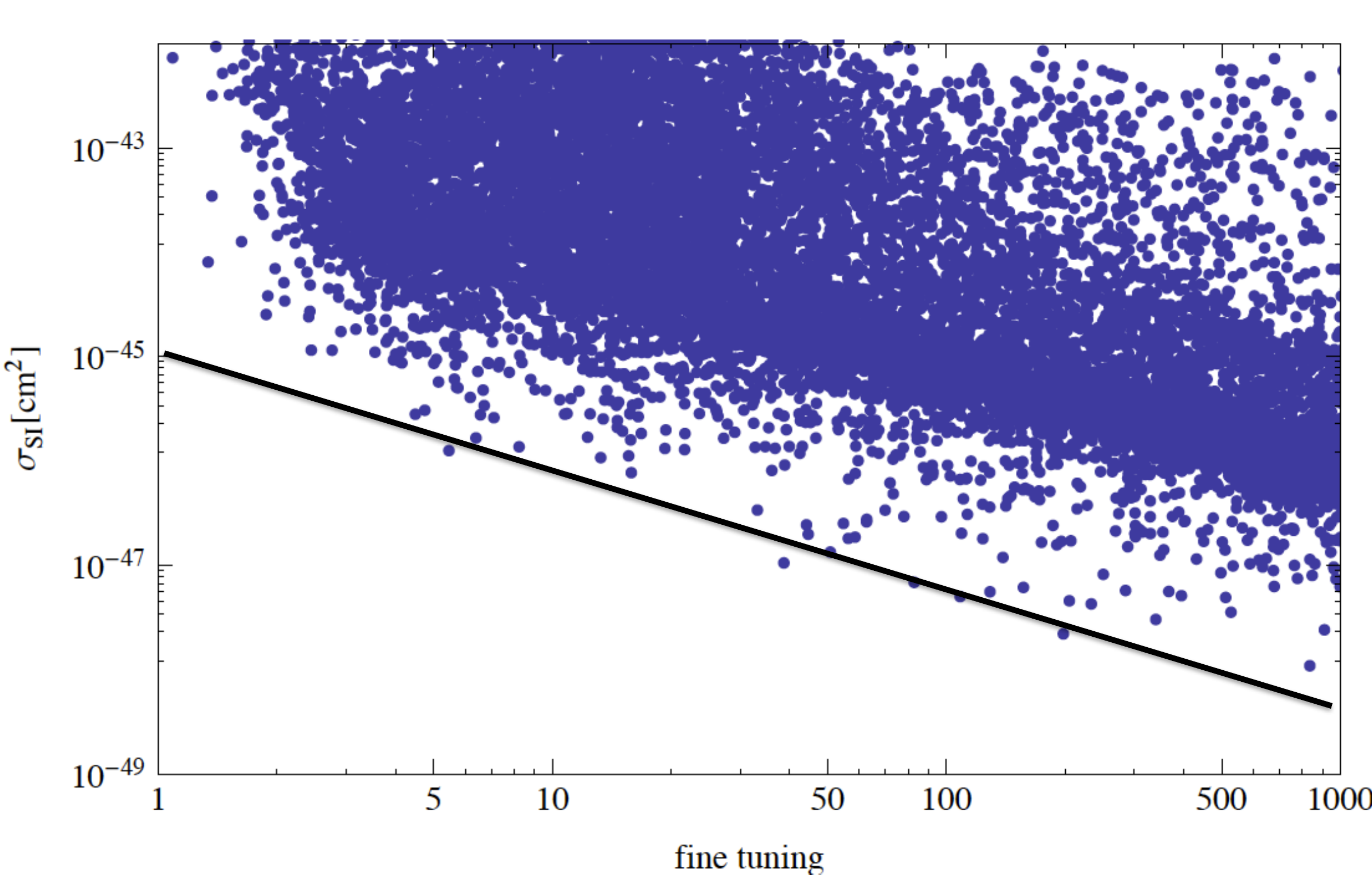}
\includegraphics[width=3.0in,height=2.1in]{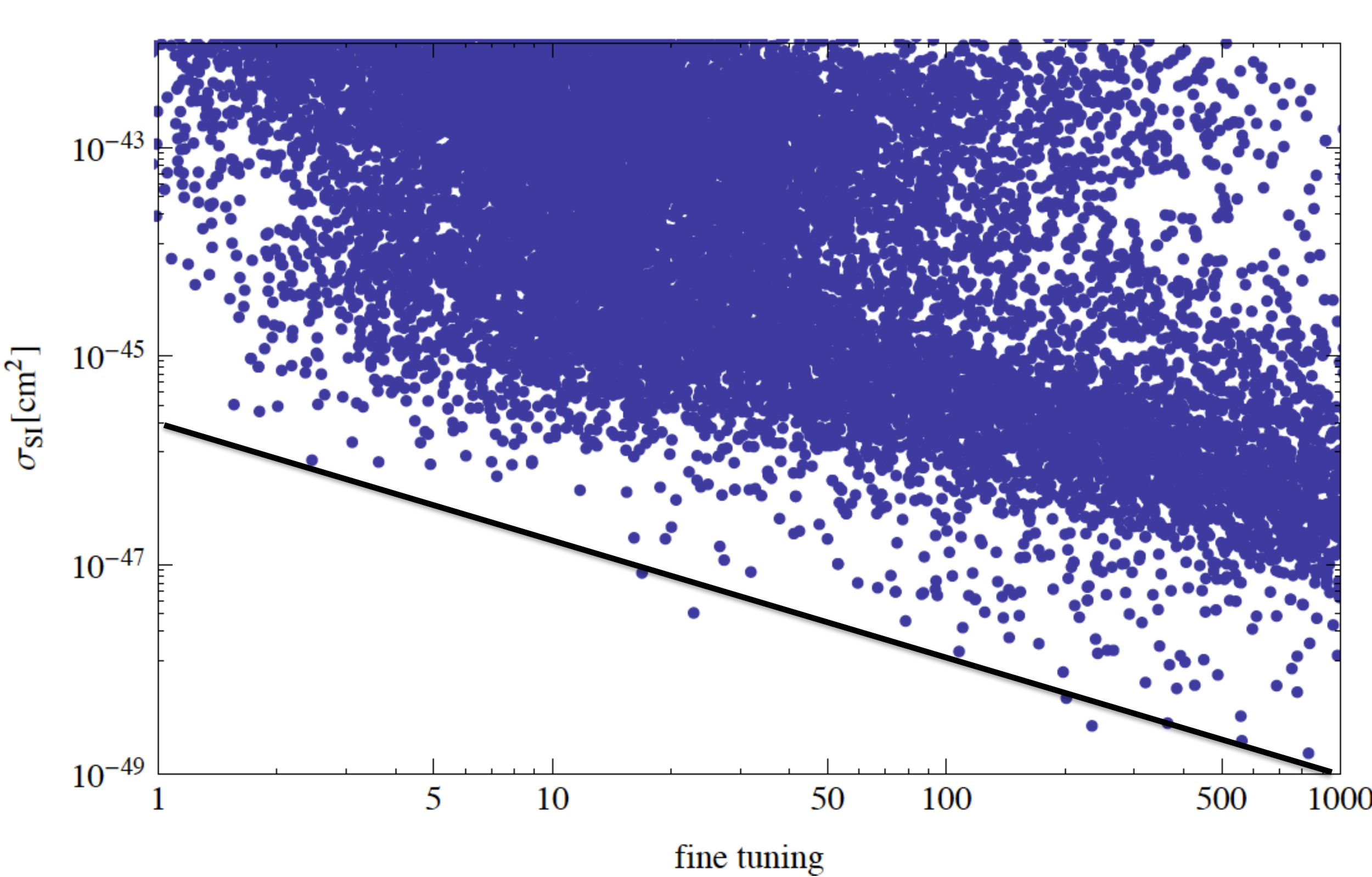}
\caption{Direct detection cross section vs. EWSB fine-tuning, in the NMSSM (left) and $\lambda$-SUSY (right). The lines represent approximate lower bounds.}
\label{fig:csft}
\end{figure}

It is instructive to again study the LSP-Higgs coupling, which feeds into the direct detection cross-section. In these models, this coupling is given by
\begin{eqnarray}
\label{coupling}
g(h_i\tilde{\chi}^0_1\tilde{\chi}^0_1) &=& \frac{\lambda}{\sqrt{2}}(s_{H_d}n_{\tilde{H}_u} n_{\tilde{S}} + s_{H_u}n_{\tilde{H}_d}n_{\tilde{S}}+s_Sn_{\tilde{H}_u}n_{\tilde{H}_d})-\frac{\kappa}{\sqrt{2}}\,s_Sn_{\tilde{S}}n_{\tilde{S}}\nonumber\\
&&+\frac{g_1}{2}(s_{H_d}n_{\tilde{B}}n_{\tilde{H}_d}-s_{H_u}n_{\tilde{B}}n_{\tilde{H}_u})
-\frac{g_2}{2}(s_{H_d}n_{\tilde{W_0}}n_{\tilde{H}_d}-s_{H_u}n_{\tilde{W_0}}n_{\tilde{H}_u})\,,
\end{eqnarray} 
where $s_\alpha$ and $n_\beta$ denote the relevant components of the $i$-th Higgs mass eigenstate ($i=1\ldots 3$) and the lightest neutralino, respectively. The final line represents the MSSM coupling and, as in the MSSM, this contribution leads to a large cross section if the LSP is gaugino-higgsino. In addition, in first term, the $\lambda$ coupling is required to be large in order to naturally give a 126 GeV Higgs, hence a singlino-higgsino LSP also results in a large cross section. A pure Higgsino is again required to be heavy from relic density considerations, and therefore fine-tuned. Higgsino-singlino mixing occurs through the $\lambda$ coupling and is therefore significant, hence a pure singlino requires the Higgsinos to be very heavy to suppress this mixing, which again leads to fine-tuning. Gauginos and gaugino-singlino mixtures are also fine-tuned from the same requirement of needing to raise the Higgsino masses. In this way, all possibilities lead to either large cross sections or large fine-tuning. This correlation is demonstrated in scans for both the NMSSM and $\lambda$-SUSY in Figure \ref{fig:csft}; for details of the scan, the reader is referred to \cite{ftsusy2}. As in the MSSM, points with accidental cancellations in cross section have been discarded.

\begin{figure}[h]
\centering
\includegraphics[width=3.0in,height=2.1in]{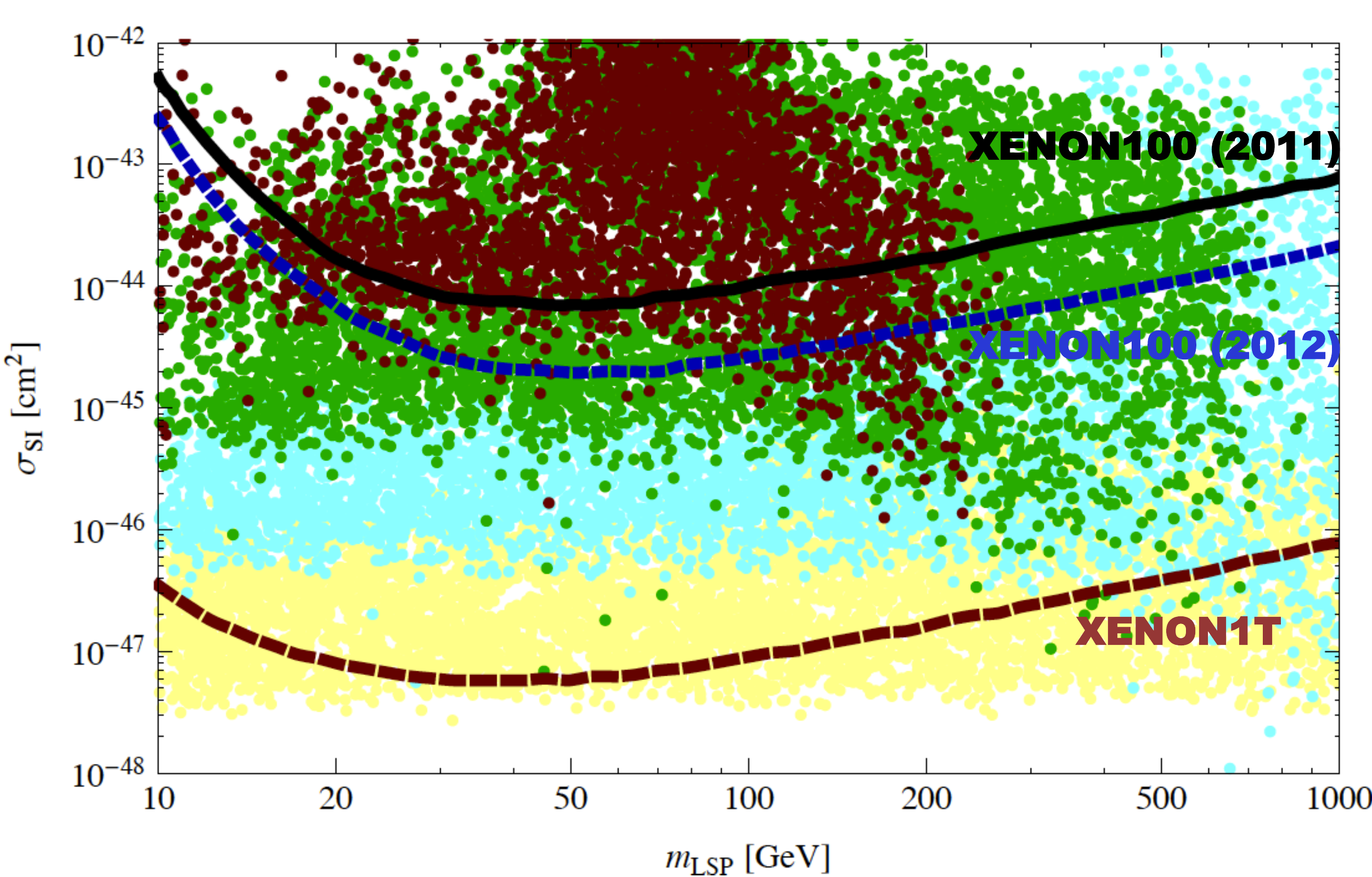}
\includegraphics[width=3.0in,height=2.1in]{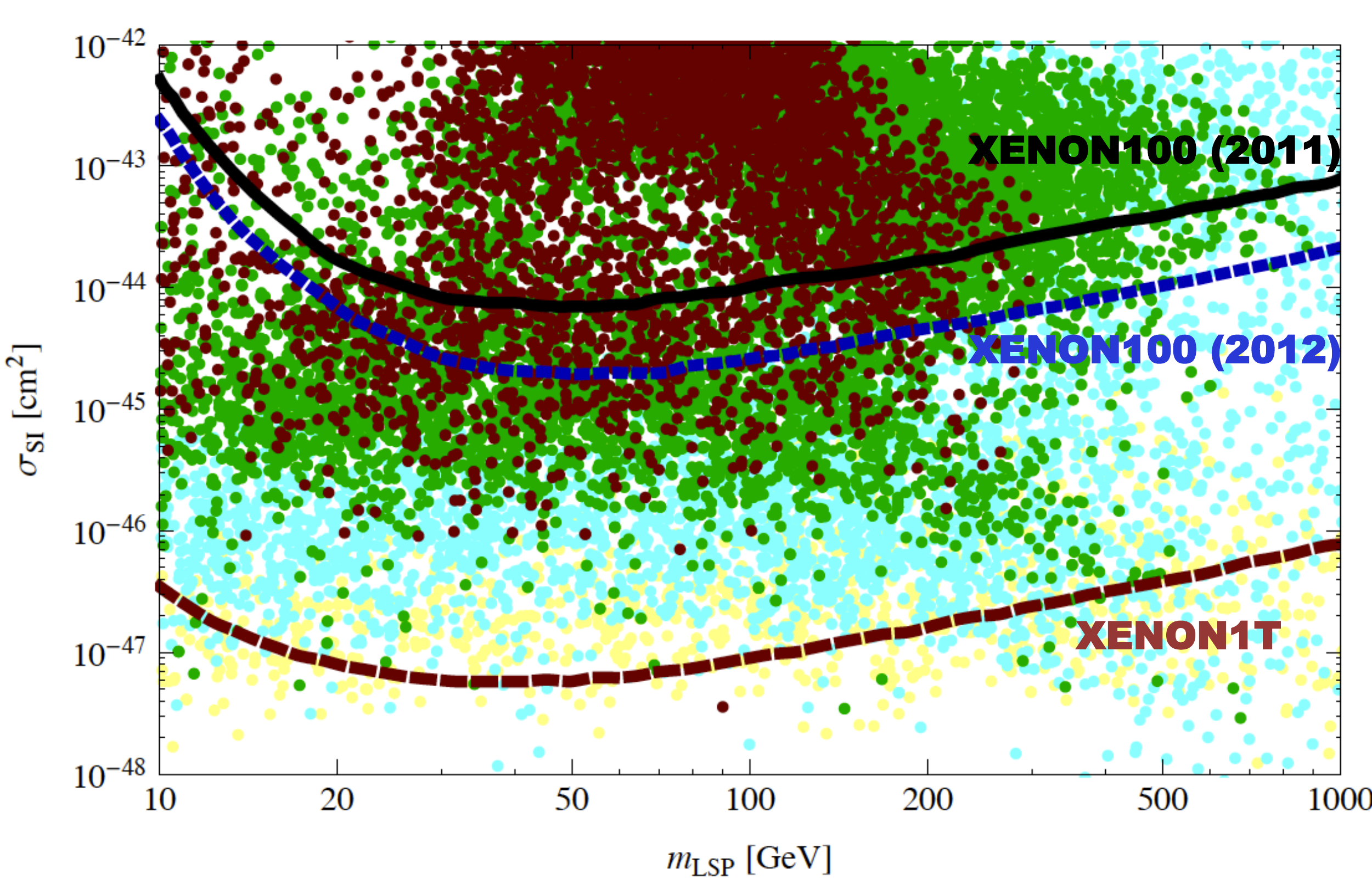}
\caption{Direct detection cross section vs. LSP mass, in the NMSSM (left) and $\lambda$-SUSY (right). Red, green, cyan and yellow points correspond to EWSB fine-tuning in the intervals $(0, 10)$; $[10, 100)$; $[100,1000)$, and $>1000$, respectively.  Points with accidental cancellations are discarded. Lines denote XENON bounds and projections}
\label{fig:csft_mass}
\end{figure}

Figure \ref{fig:csft_mass} plots the cross section as a function of the dark matter mass against the current and projected XENON bounds for both models. In both Figure \ref{fig:csft} and \ref{fig:csft_mass}, it is seen that the amount of fine-tuning is generally lower for $\lambda$-SUSY compared to the MSSM; this is one of the attractive features of $\lambda$-SUSY, coming from the large value of $\lambda$, which improves EWSB fine-tuning by a factor of roughly $(\lambda/g)^2$\footnote{See \cite{ftsusy2} and references therein for more detail. On the other hand, large $\lambda$ now incurs fine-tuning with respect to the Higgs data, see \cite{lsusyhiggs} and references therein.}. As a consequence of this, while the most natural points (fine-tuning\textless 10) in the NMSSM are in tension with the XENON100 constraint, parts of this region are still below the XENON100 bound in $\lambda$-SUSY. However, the XENON1T projected bound should probe both models to percent-level fine-tuning.

\section{Conclusions}
The Fermi 130 GeV gamma ray signal is inconsistent with neutralino annihilation into two photons; however, internal bremsstrahlung from bino dark matter with nearly degenerate sleptons was found to provide a viable mechanism to explain the signal, with a very efficient suppression of the problematic continuum, albeit with all parameters stretched to their very limits. On the direct detection front, low direct detection cross sections were found to be correlated with greater fine-tuning in EWSB (at tree-level) in the MSSM, NMSSM, and $\lambda$-SUSY. The most natural points (fine-tuning \textless 10) in the MSSM and the NMSSM are in some tension with the XENON100 bound already, whereas $\lambda$-SUSY remains more natural. XENON1T is projected to probe all models down to percent level tuning. Neutralino dark matter, therefore, continues to proceed into a rapidly evolving era. 


\begin{theacknowledgments}
The author would like to thank CETUP* (Center for Theoretical Underground Physics and Related Areas), supported by the US Department of Energy under Grant No. DE-SC0010137 and by the US National Science Foundation under Grant No. PHY-1342611, for its hospitality and partial support during the 2013 Summer Program, where these results were presented. The papers \cite{ftmssm} and \cite{ftsusy2} presented at the workshop and in this writing were written in collaboration with Maxim Perelstein. 
\end{theacknowledgments}



\bibliographystyle{aipproc}   


\begin{thebibliography}{9}

 \bibitem{HPR}
  L.~J.~Hall, D.~Pinner and J.~T.~Ruderman,
 {\it ``A Natural SUSY Higgs Near 126 GeV,''}
  JHEP {\bf 1204}, 131 (2012)
  [arXiv:1112.2703 [hep-ph]].

\bibitem{xenon100paper}
  E.~Aprile {\it et al.}  [XENON100 Collaboration],
 {\it ``Dark Matter Results from 100 Live Days of XENON100 Data,''}
  Phys.\ Rev.\ Lett.\  {\bf 107}, 131302 (2011)
  [arXiv:1104.2549 [astro-ph.CO]].

  \bibitem{xenon2012} 
    E.~Aprile  {\it et al.}  [XENON100 Collaboration],
 {\it ``Dark Matter Results from 225 Live Days of XENON100 Data,''}
  arXiv:1207.5988 [astro-ph.CO].
  
  \bibitem{130weniger}
    C.~Weniger,
  ``A Tentative Gamma-Ray Line from Dark Matter Annihilation at the Fermi Large Area Telescope,''
  JCAP {\bf 1208}, 007 (2012)
  [arXiv:1204.2797 [hep-ph]];
  
  \bibitem{130fermi} 
  M.~Su and D.~P.~Finkbeiner,
  ``Strong Evidence for Gamma-ray Line Emission from the Inner Galaxy,''
  arXiv:1206.1616 [astro-ph.HE];\\
  E.~Tempel, A.~Hektor and M.~Raidal,
  {\it``Fermi 130 GeV gamma-ray excess and dark matter annihilation in sub-haloes and in the Galactic centre,''}
  arXiv:1205.1045 [hep-ph];\\
  T.~Bringmann, X.~Huang, A.~Ibarra, S.~Vogl and C.~Weniger,
  ``Fermi LAT Search for Internal Bremsstrahlung Signatures from Dark Matter Annihilation,''
  JCAP {\bf 1207}, 054 (2012)
  [arXiv:1203.1312 [hep-ph]].
  
    \bibitem{continuum} 
  T.~Cohen, M.~Lisanti, T.~R.~Slatyer and J.~G.~Wacker,
  ``Illuminating the 130 GeV Gamma Line with Continuum Photons,''
  arXiv:1207.0800 [hep-ph];
  
  \bibitem{continuum2}
  W.~Buchmuller and M.~Garny,
  ``Decaying vs Annihilating Dark Matter in Light of a Tentative Gamma-Ray Line,''
  JCAP {\bf 1208}, 035 (2012)
  [arXiv:1206.7056 [hep-ph]];\\
  I.~Cholis, M.~Tavakoli and P.~Ullio,
  ``Searching for the continuum spectrum photons correlated to the 130 GeV gamma-ray line,''
  arXiv:1207.1468 [hep-ph].
  
   \bibitem{ib} 
  T.~Bringmann, L.~Bergstrom and J.~Edsjo,
  ``New Gamma-Ray Contributions to Supersymmetric Dark Matter Annihilation,''
  JHEP {\bf 0801}, 049 (2008)
  [arXiv:0710.3169 [hep-ph]];\\
  L.~Bergstrom, T.~Bringmann, M.~Eriksson and M.~Gustafsson,
  ``Gamma rays from heavy neutralino dark matter,''
  Phys.\ Rev.\ Lett.\  {\bf 95}, 241301 (2005)
  [hep-ph/0507229];\\
   A.~Birkedal, K.~T.~Matchev, M.~Perelstein and A.~Spray,
  ``Robust gamma ray signature of WIMP dark matter,''
  hep-ph/0507194.
  
    \bibitem{130susy} 
  B.~Shakya,
  {\it ``A 130 GeV Gamma Ray Signal from Supersymmetry,''}
  Phys.\ Dark Univ.\  {\bf 2}, 83 (2013)
  [arXiv:1209.2427 [hep-ph]].
  
   \bibitem{tcreview} 
  T.~Bringmann and C.~Weniger,
  ``Gamma Ray Signals from Dark Matter: Concepts, Status and Prospects,''
  arXiv:1208.5481 [hep-ph].
  
  \bibitem{positrons} 
  M.~Perelstein and B.~Shakya,
  ``Remarks on calculation of positron flux from galactic dark matter,''
  Phys.\ Rev.\ D {\bf 82}, 043505 (2010)
  [arXiv:1002.4588 [astro-ph.HE]];\\
  P.~Meade, M.~Papucci, A.~Strumia and T.~Volansky,
  ``Dark Matter Interpretations of the e+- Excesses after FERMI,''
  Nucl.\ Phys.\ B {\bf 831}, 178 (2010)
  [arXiv:0905.0480 [hep-ph]].
  
  \bibitem{ftmssm}
  M.~Perelstein and B.~Shakya,
 {\it ``Fine-Tuning Implications of Direct Dark Matter Searches in the MSSM,''}
  JHEP {\bf 1110}, 142 (2011)
  [arXiv:1107.5048 [hep-ph]].
  
    \bibitem{ftsusy2}
  M.~Perelstein and B.~Shakya,
  ``XENON100 Implications for Naturalness in the MSSM, NMSSM and lambda-SUSY,''
  arXiv:1208.0833 [hep-ph].
  
  \bibitem{welltemp}
  N.~Arkani-Hamed, A.~Delgado and G.~F.~Giudice,
  {\it ``The Well-tempered neutralino,''}
  Nucl.\ Phys.\  B {\bf 741}, 108 (2006)
  [arXiv:hep-ph/0601041].
  
   \bibitem{ftmssm2} 
   S.~Amsel, K.~Freese and P.~Sandick,
{\it  ``Probing EWSB Naturalness in Unified SUSY Models with Dark Matter,''}
  JHEP {\bf 1111}, 110 (2011)
  [arXiv:1108.0448 [hep-ph]];
  
  \bibitem{lsusyhiggs} 
  M.~Farina, M.~Perelstein and B.~Shakya,
  {\it ``Higgs Couplings and Naturalness in lambda-SUSY,''}
  arXiv:1310.0459 [hep-ph].

\end{thebibliography}



\end{document}